\documentclass[twocolumn,showpacs,preprintnumbers,amsmath,amssymb]{revtex4}

\usepackage{graphicx}
\usepackage{amsmath,amssymb}
\usepackage{dcolumn}
\usepackage{bm}
\usepackage{color,ulem}

\usepackage{subfigure}
\newlength{\subfigwidth}
\newlength{\subfigcolsep}
\setkeys{Gin}{width=\subfigwidth}

\begin{document}   
\title{
Inelastic collapse in one-dimensional driven systems under gravity
}

\author{Jun'ichi Wakou$^{1,2}$}
 \email{wakou@cc.miyakonojo-nct.ac.jp}
\author{Hiroyuki Kitagishi$^{2}$}
\author{Takahiro Sakaue$^{2}$}
\author{Hiizu Nakanishi$^{2}$}
 \email{nakanisi@phys.kyushu-u.ac.jp}

\affiliation{
$^{1}$Miyakonojo National College of Technology, Miyakonojo-shi, Miyazaki, 885-8567, Japan\\
$^{2}$Department of Physics, Kyushu University 33, Fukuoka 812-8581, Japan
}

\date{\today}

\begin{abstract}
We study the inelastic collapse in the one-dimensional $N$-particle
systems in the situation where the system is driven from below under the
gravity.
We investigate the hard-sphere limit of
the inelastic soft-sphere systems by numerical simulations
to find how the collision rate per particle
$n_{\rm coll}$ increases as a function of the elastic constant of
the sphere $k$
when the restitution coefficient $e$  is kept constant.
For the systems with large enough $N\agt 20$, we find three regimes in
$e$ depending on the behavior of $n_{\rm
coll}$ in the hard-sphere limit:
(i) {\it uncollapsing regime} for $1\ge e>e_{c1}$, where $n_{\rm coll}$
converges to a finite value,
(ii) {\it logarithmically collapsing regime} for $e_{c1}> e > e_{c2}$,
where $n_{\rm coll}$ diverges
as $n_{\rm coll}\sim \log k$, and
(iii) {\it power-law collapsing regime} for $e_{c2}>e>0$, where $n_{\rm
coll}$ diverges
as $n_{\rm coll} \sim k^\alpha$ with an
exponent $\alpha$ that depends on $N$.
The power-law collapsing regime shrinks as $N$ decreases and seems not
to exist for the system with $N=3$ while, for large $N$, the size of the
uncollapsing and the logarithmically collapsing regime decreases as
$e_{c1} \simeq 1-2.6/N$ and $e_{c2} \simeq 1-3.0/N$ .
We demonstrate that this difference between large and small
systems exists already in the inelastic collapse without the
external drive and the gravity.
\end{abstract}


\maketitle

\section{Introduction}

The inelastic hard-sphere system is one of the
simplest models of granular media.  It
consists of rigid spheres that interact with each
other only through instantaneous inelastic collisions.
Minimum ingredients of the granular systems are taken in this system,
for which the efficient event-driven algorithms have been developed for
molecular dynamics simulations~\cite{Rapaport-1980,Luding-2004,Isobe-1999} as well as sophisticated kinetic theories
for analytical study (see, for example, Ref.~\cite{BrilliantovPoschel-2004}).

With this idealization of the granular media, however, it has
been known that infinite number of collisions among a finite
number of particles can occur in a finite length of time. This
phenomenon is called {\it inelastic collapse}~\cite{BernuMazighi-1990,McNamaraYoung-1992}.  Process of
collisions involved in the inelastic collapse has been studied for
one-dimensional (1-d)~\cite{BernuMazighi-1990,McNamaraYoung-1992} and
two-dimensional (2-d) systems~\cite{McNamaraYoung-1994,McNamaraYoung-1996,ZhouKadanoff-1996,SchorghoferZhou-1996,AlamHrenya-2001},
and the conditions for the inelastic collapse have been obtained
in some situations~\cite{BernuMazighi-1990,McNamaraYoung-1992,McNamaraYoung-1994,McNamaraYoung-1996,ZhouKadanoff-1996,SchorghoferZhou-1996}.

One of the simplest cases is the freely cooling granular gas, in
which the inelastic hard-sphere system develops freely without any
external forces~\cite{McNamaraYoung-1994,SchorghoferZhou-1996}.  In
the 2-d systems, it has been shown that the particles that
partake in the inelastic collapse form a string like linear structure
for the case of frictionless particles~\cite{McNamaraYoung-1994},
while they form a string like zigzag pattern for the case of the
frictional particles~\cite{SchorghoferZhou-1996}.
Another simple case is a simple shear flow, where the collapsing
particles have been shown to form a linear string structure
typically oriented along the direction
$45^\circ$ from the flow direction~\cite{AlamHrenya-2001}.


In the hard-sphere idealization, once the inelastic collapse occurs, the
system cannot proceed further without additional assumptions for the
dynamics, such as those in the contact
dynamics~\cite{JeanMoreau-1992,Moreau-1994}.  A simple way to escape
from this difficulty is to suppress the inelastic collapse by employing
the velocity dependent restitution coefficient that goes to one as the
colliding velocity goes to
zero~\cite{Goldmanetal-1998,Bizonetal-1998}\footnote{ This is actually
not completely fictitious model because the dissipation often decreases
for low velocity collisions in real systems~\cite{Goldsmith-1960}.}.  In
actual systems with finite rigidity, the inelastic collapse should never
occur.

Although the inelastic collapse is a singular behavior in the idealized
system of the infinitely hard spheres with a constant restitution
coefficient, its relevance to some physical behaviors has been
suggested.  In flowing configurations, it has been demonstrated by
numerical simulations that there exists strong correlation between the
force chain network and the chain like structure formed by particles
that collide repeatedly with each other in the hopper
flow~\cite{FergusonChakraborty-2006}.  The inelastic collapse has been
also discussed in connection with the formation of correlation in the
shear flow~\cite{LoisLemaitreCarlson-2007}.

In order to study how the inelastic collapse affects system
behaviors in physical situations, it is natural to investigate the
soft sphere system with finite rigidity and see how the inelastic
collapse appears in the hard-sphere limit.  If you take, however, a
simple limit of the infinite elastic constant with finite dissipation
parameters, the resulting restitution coefficient tends to one,
therefore the inelastic collapse does not occur.  Thus the pertinent
hard sphere limit for this purpose is the limit of infinite elastic
constant with keeping the resulting restitution coefficient constant by
making the dissipation parameter infinite.  
%

Mitarai and Nakanishi studied such  limit by examining the limiting
behavior of the collision rate $n_{\rm coll}$ for the 2-d gravitational
flow~\cite{MitaraiNakanishi-2003}.  The hard-sphere limit was taken as
the limit of the infinite elastic constant $k$ with the restitution
coefficient $e$ being kept constant.  They found that $n_{\rm coll}$
converges to a finite value in the collisional flow regime, while it
diverges as $n_{\rm coll}\sim k^{\alpha}$ as $k\to\infty$ in the
frictional flow regime.  The exponent $\alpha$ was estimated to be about
$0.4$ in their case, i.e. the 2-d gravitational flow on a flat slope
with ten layers of particles and the restitution coefficient $e=0.7$.
More recently, Brewster et al. studied the three-dimensional
gravitational flow and obtained $\alpha \simeq 0.25$ for the system with
90$\sim$100 layers of particles on a rough slope and
$e=0.88$~\cite{BrewsterSilbertGrestLevine-2008}.
Although the divergence of collision rate implies emergence of inelastic
collapse in the hard-sphere limit, a simple consideration on
exponentially decreasing collision time interval would give the
logarithmic divergence, and the mechanism for the power law divergence
has not been understood yet.

Motivated by these findings of the power law divergence in the
gravitational slope flow, in this paper we take a closer look at the
problem in an even simpler system, namely, a 1-d inelastic particle
system under the gravity with an external excitation from a bottom of
the system.  The external excitation at the bottom is supposed to mimic
the excitation by random collisions of particles with the slope in the
gravitational flow, and our 1-d system is intended to capture only the
particle motion perpendicular to the slope.
By numerical simulations,
we will show that even this simple model exhibits the power law
divergence of the collision rate, $n_{\rm coll}\sim k^{\alpha}$.
%

This paper is organized as follows.  In Sec.~II, we begin by
introducing our model and describe a method to study
the hard-sphere limit of soft spheres in our simulations.
The system with $N=3$ is analyzed to show
the logarithmic behavior $n_{\rm coll}\sim \log k$ in the hard sphere limit.
In Sec.~III, after describing the simulation procedure, first we present
the simulation results for the systems with small number of particles
($N=3\sim 6$); Only the logarithmic behavior in the hard sphere limit is
observed for $N=3$ while the power law divergence regime appears for the
larger system in the smaller $e$ region.  Then we show the simulation
results of systems with large number of particles ($N\agt 20$) and
demonstrate that there exist the three distinct regimes for the limiting
behavior of $n_{\rm coll}$ in $e$.  We discuss the origin of these
limiting behaviors based on the simulation results of inelastic collapse
in the 1-d free space.  Summary and conclusion are given in Sec.~IV.

\section{One-dimensional model of granular flow}
\subsection{Model}

\begin{figure}[b]
\begin{center}
\includegraphics[width=\hsize,clip]{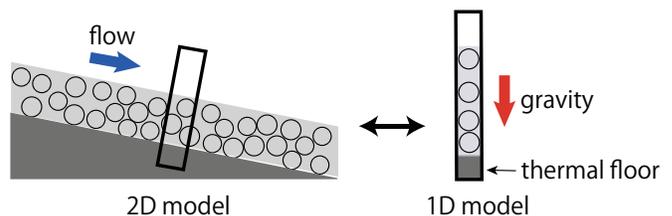}
\caption{A schematic representation of relation between the
granular slope flow and the 1-d driven system.}  \label{1dmodel}
\end{center}
\end{figure}

We consider the 1-d model by focusing
the particle motion only perpendicular to the
slope
(see Fig.~\ref{1dmodel}).
The particles are allowed to move only along the $z$-axis under
the influence of gravity
and the lowest particle is excited by the bottom floor.

Let us consider $N$ identical particles with mass $m$ and diameter $d$.
The particles are numbered from the bottom starting with $i=1$, and can
interact only with their adjacent particles through the soft-sphere
interaction.
The excitation by the random collision with the slope is represented by
the thermal floor located at the bottom $z=0$.
When the lowest particle ($i=1$)
collides with the bottom, it comes off with a random velocity $v$ by the Maxwell-Boltzmann distribution
\begin{eqnarray}
p(v)= \frac{mv}{k_{\rm B} T_0}\exp\left(-\frac{mv^2}{2k_{\rm B} T_0}\right),
\end{eqnarray}
where $T_0$ is temperature of the thermal floor and $k_{\rm B}$ is the
Boltzmann constant.

The interaction between soft spheres is given by the so-called {\it
spring-dashpot} model~\cite{Duran-2000,Luding-2004}.  Let  $z_i$ and $v_i$
 denote the coordinate and the velocity of
particle $i$, respectively, then the overlap
 between the adjoining two particles $i$ and $i+1$ is given by
$x_{i,i+1}\equiv d-(z_{i+1}-z_{i})$.  The relative velocity between $i$
and $i+1$ is denoted as
$v_{i,i+1}\equiv v_{i}-v_{i+1}={d x_{i,i+1}}/{dt}$.
Then, the force $f_{i,i+1}$ exerted
on particle $i$ by particle $i+1$ is given by
\begin{eqnarray}
f_{i,i+1}
=\left\{
\begin{array}{@{\,}ll}
-k x_{i,i+1}-Dm v_{i,i+1} & \mbox{(for $x_{i,i+1}>0$)},\\
0 & \mbox{(for $x_{i,i+1}\le 0$)}.
\end{array}
\right.
\label{force-0}
\end{eqnarray}
The first term of Eq.~(\ref{force-0}) represents the elastic force
by the Hookean law with the elastic constant $k$. The second
term denotes the dissipative force proportional to the relative
velocity $v_{i,i+1}$, where $D$ is the damping constant.  The force
acting on the particle $i+1$ by the particle $i$ is given by
$f_{i+1,i}=-f_{i,i+1}$.
Note that the dissipative force is discontinuous at $x_{i,i+1}=0$.

The equation of motion for the particle $i$ is then given by
\begin{eqnarray}
m\frac{d v_i}{dt} = -mg +f_{i,i+1}-f_{i-1,i},
\end{eqnarray}
where $g$ is the gravitational acceleration.  For our linear force law
of Eq.~(\ref{force-0}), the duration time of contact for a binary
collision $\tau_{c}$ is constant and given by
\begin{eqnarray}
 \tau_{c}=\frac{\pi}{\sqrt{(2k/m)-D^2}}.
\label{tauc}
\end{eqnarray}

\subsection{Hard-sphere limit}

For the linear force law in Eq.~(\ref{force-0}),
the restitution coefficient $e$ of a binary collision is given by
\begin{eqnarray}
e\equiv
-\frac{\left. v_{i,i+1}\right|_{t=\tau_{c}}}
                      {\left. v_{i,i+1}\right|_{t=0}}
=\exp\left(-D \tau_{c}\right),
\end{eqnarray}
using the duration time $\tau_{c}$ of Eq.(\ref{tauc}).
By solving this for $D$, we obtain
\begin{eqnarray}
 D=\sqrt{\frac{2k}{m}\cdot\frac{(\ln e)^2}{\pi^2+(\ln e)^2}}.
\label{Dd}
\end{eqnarray}
The hard-sphere limit is defined as
the limit of $k\to \infty$ with keeping $e$ constant.
Thus, in this limit $D$ diverges as $D\propto k^{1/2}$ and $\tau_{c}$ goes to zero as $\tau_{c}\propto k^{-1/2}$.

\subsection{Inelastic collapse}

Bernu and Mazighi~\cite{BernuMazighi-1990} have studied $N$ inelastic
hard spheres thrown against a wall in 1-d system
and showed the inelastic collapse can occur if $e$ is less than a critical
value $e_{c}^{\rm wall}(N)$.  They gave an analytical expression for
$e_{c}^{\rm wall}(N)$ using the independent collision wave (ICW) model:
\begin{eqnarray}
  e_{c}^{\rm wall}(N)=
    \tan^2\left(\frac{\pi}{4}\left(1-\frac{1}{N}\right)\right).
\label{icw}
\end{eqnarray}
This is exact for the case of $N=2$ but is an approximation for $N>2$
because the model ignores interaction between
collision waves. Using another model called the cushion model,
McNamara and Young~\cite{McNamaraYoung-1992}
have obtained an estimate for
the minimum number of particles $N_{c}^{\rm wall}$ that is
required for collapse when the restitution coefficient is $e$:
 \begin{eqnarray}
  N_{c}^{\rm wall}(e)=\frac{\ln(4/(1-e))}{1-e}.
 \end{eqnarray}
This result becomes exact in the limit $e\to 1$.  Comparison between the
ICW model and the cushion model has been discussed in
Refs.~\cite{McNamaraYoung-1992,McNamara-2002}, and the
numerical simulations show that the former is more
accurate for $N<15$ while the latter is better for $N>15$.
In the large $N$ limit, both of the models give $e_{c}^{\rm wall}\to 1$, but the asymptotic forms are
\begin{eqnarray}
 e_{c, {\rm ICW}}^{\rm wall}\approx 1-\frac{\pi}{N},
\label{ec_icw_wall}
\end{eqnarray}
for the ICW model and
\begin{eqnarray}
 e_{c, {\rm cushion}}^{\rm wall}\approx 1-\frac{1}{N}\ln(4N)\left(1-\frac{\ln\ln 4N}{\ln 4N}\right),
\label{ec_cushion}
\end{eqnarray}
for the cushion model.

\begin{figure}[b]
\begin{center}
\includegraphics[width=0.5\hsize,clip]{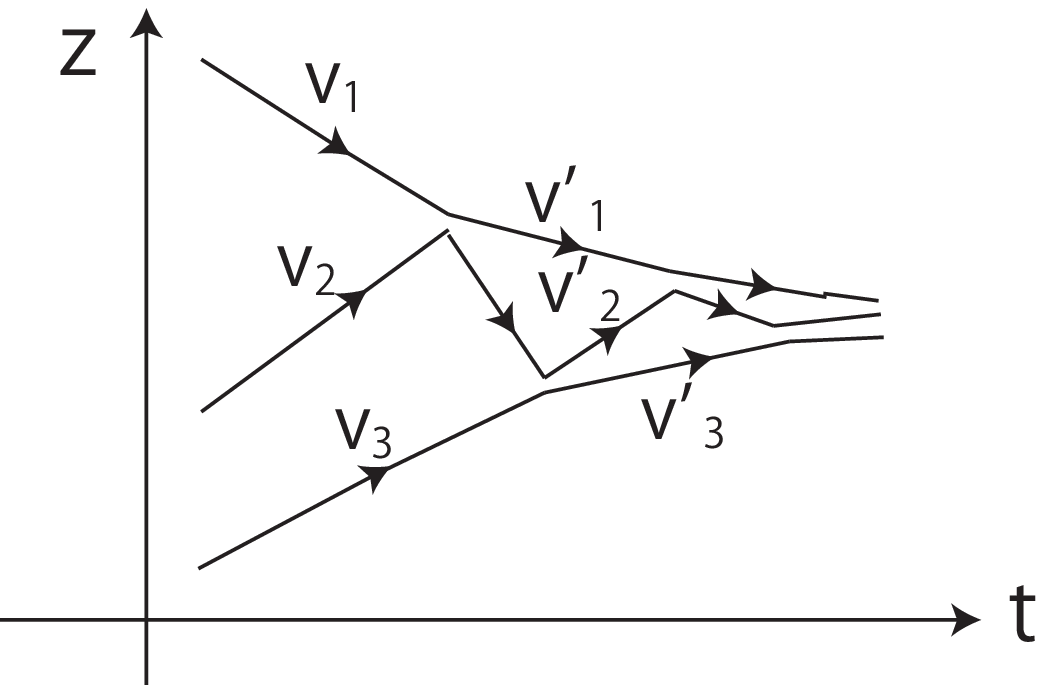}
\caption{
Schematic diagram for particle motion in the
three body inelastic collapse in the one dimensional free space.
}
\label{threebodycollapse}
\end{center}
\end{figure}

The inelastic collapse can also occur in free space in
1-d system if $e$ is less than a critical value $e_{c}(N)$.  A
schematic picture of the three-body inelastic collapse is given in
Fig.~\ref{threebodycollapse}.
McNamara and Young~\cite{McNamaraYoung-1992} have shown that
the three-body inelastic collapse can occur if $e<e_{c}(3)\equiv 7-4\sqrt{3}$,
by using the $3\times 3$ matrix $\cal{M}$ that relates the final velocities
$\bm{v}'=(v'_1,v'_2,v'_3)$ of the three particles after two successive
collisions with their initial velocities $\bm{v}=(v_1,v_2,v_3)$
as $\bm{ v}'={\cal M} \bm{v}$.

For $N=4$, the critical value can be evaluated as
$e_{c}(4)=e_{c}^{\rm wall}(2)$ because of the symmetry in the order
of collisions.
If such symmetry in the order of collision process is
assumed for $N>4$, one may obtain the relation
 \begin{eqnarray}
  e_{c}(2N)=e_{c}^{\rm wall}(N),
 \label{symmetry}
\end{eqnarray}
but numerical simulations have shown that the relation Eq.~(\ref{symmetry}) is not valid
for large $N$~\cite{McNamara-2002}.

\subsection{Asymptotic analysis for $N=3$}
In this subsection, we examine the asymptotic behavior of the total
number of collisions $n_{\rm tot}$ in a single collapsing event in the
hard-sphere limit for the cases of $N=3$ and show that $n_{\rm tot}\sim
\log k$ in the $k\to \infty$ limit.

The system with $N=3$ is the smallest one where the inelastic collapse can occur,
since the floor provides a thermal drive, thus the inelastic collapse can happen only
in sequence of collisions among particles.
Then, we can argue the behavior of $n_{\rm tot}$ by considering a
collision process of the three-body inelastic collapse in
the hard-sphere limit as shown in Fig.~\ref{threebodycollapse}.
In this case,
the time between the $(n-1)$th collision
and $n$th collision, $t_{12}^{(n)}$,
between the same pair of particles, say 1 and 2,
behaves as
\begin{eqnarray}
t_{12}^{(n)}\approx  q t_{12}^{(n-1)}\approx q^n t_{12}^{(0)},
\label{deltat12}
\end{eqnarray}
where $q$ is a constant smaller than
unity (See Appendix A).

Now, let us discuss the case of the soft spheres with a finite $k$
and $e<e_{c}(3)$.  In this case, initial binary collisions can follow a
sequence similar to the inelastic collapse, but eventually all of the
three particles are in contact after a finite number of collisions, and
then fly away from each other with very small relative velocities.
We estimate the total number $n_{\rm tot}$ of collisions before all three
particles are in contact at the same time. Similar
estimation has been done for the case of a single inelastic soft sphere
bouncing on a floor to show $n_{\rm tot}\sim
\log k$ as $k\to \infty$~\cite{MitaraiNakanishi-2003}.
The three-body collapse like collision process shows
essentially the same behavior as we will show in the following.

First of all, the collision interval $t_{12}^{(n)}$ for the case of the soft-sphere system is given by
\begin{eqnarray}
t_{12}^{(n)}&\approx&
q t_{12}^{(n-1)}+\Delta t_{12},
\label{t12ss}
\end{eqnarray}
with the correction term $\Delta t_{12}$ in comparison with Eq.~(\ref{deltat12})
because the collision duration $\tau_{c}$ is finite.
It can be shown (see Appendix B) that
\begin{eqnarray}
 \Delta t_{12}=-f\tau_{c},
\label{deltat12ss}
\end{eqnarray}
where $f$ is positive and a function of $e$.
Substituting Eq.~(\ref{deltat12ss}) into Eq.~(\ref{t12ss}), we obtain
\begin{eqnarray}
t_{12}^{(n)}&\approx&
q t_{12}^{(n-1)}-f\tau_{c}
\nonumber\\
&\approx &
q^{n} t_{12}^{(0)}-\sum_{i=0}^{n-1}q^{i} f\tau_{c}
\nonumber\\
& = &
q^{n} t_{12}^{(0)}-\frac{1-q^{n}}{1-q} f\tau_{c}
\label{deltat12ss_2}
\end{eqnarray}
The number of collisions $n_{\rm tot}$ before
all three particles are in contact at the same time is given by the smallest $n$
that satisfies the condition
$t_{12}^{(n+1)}\le 2\tau_{c}$,
because two successive collisions between particles 1 and 2
cannot be shorter than the twice of the duration time.
Thus, by requiring the relation
\begin{eqnarray}
 t_{12}^{(n_{\rm tot})}\simeq 2\tau_{c},
\end{eqnarray}
for $n_{\rm tot}\gg 1$ and substituting Eq.~(\ref{deltat12ss_2}), we obtain
\begin{eqnarray}
q^{n_{\rm tot}} t_{12}^{(0)}\approx \left( 2+\frac{1-q^{n_{\rm tot}}}{1-q}f\right)\tau_{c}.
\label{ntoteq}
\end{eqnarray}
$n_{\rm tot}$ diverges and $q^{n_{\rm tot}}$ goes to zero in the limit $k\to \infty$
because $\tau_{c}\propto k^{-1/2}$ and $0\le q < 1$.
Thus,
Eq.~(\ref{ntoteq}) can be written as $q^{n_{\rm tot}} t_{12}^{(0)}\sim a k^{-1/2}$,
where $a$ is a coefficient that depends on $e$.
Therefore, we obtain $n_{\rm tot}$
\begin{eqnarray}
 n_{\rm tot}\sim -\frac{1}{2\log q}\log k + \mbox{const.}
\label{ntot-log}
\end{eqnarray}
If we {\it assume} that the frequency $r$ of such a three-body
process is independent of $k$, the collision rate is $n_{\rm coll}\sim
r\, n_{\rm tot}$ and thus $n_{\rm coll}\sim \log k$ in the hard-sphere
limit.  

\section{Simulation results}

The main quantity studied in this paper is the collision rate per
particle $n_{\rm coll}$ defined as the average number of collisions
(including collisions with the floor) per particle per unit time
for various values of parameters, $k$, $e$, $N$ and $T_0$.  We carried
out numerical simulations to investigate $n_{\rm coll}$ in the hard-sphere limit.
After describing simulation procedure, we present the results for small number of particles $3\le N\le 6$ first, and
then for large number of particles $N\agt 20$.
We find qualitative difference between the two cases.

\subsection{Simulation procedure}

Numerical simulations are performed
using the second-order Runge-Kutta method
with the time step $dt=\tau_{\rm c}/100$, where $\tau_{\rm c}$ is the duration
time of binary collision given by Eq.(\ref{tauc}).
All particles are initially placed in such a way that there is no
overlap between particles and velocities are given randomly.  After
waiting for a sufficiently long time for the system to
go through an initial transient,
we start taking data for
various quantities and their time average.

For numerical data, we employ the unit system where the particle mass
$m$, the diameter $d$, and the gravitational acceleration $g$ are
unities,
\begin{equation}
m = d = g = 1.
\end{equation}
We set temperature of the thermal floor $k_{\rm B} T_0=1$ unless
otherwise stated.  For a given set of $N$ and $e$, we measure
the collision rate $n_{\rm coll}$ for
$k=10^{5},\,10^{6},\,10^{7},\,10^{8},\,10^{9}$, and $10^{10}$.

Each collision event between two particles or between a particle
and the floor is defined by their contact.  The collisions between
particles last for some duration time and they are counted everytime
colliding particles separate, while the collisions with the floor are
assumed to be instantaneous.  The total number of collisions $N_{\rm
coll}$ includes the collisions between particles and those between a
particle and the floor, and the collision rate per particle $n_{\rm coll}$ is defined by
\begin{eqnarray}
 n_{\rm coll}=N_{\rm coll}/ (NT),
\end{eqnarray}
where $T$ is the simulation time length.
We set $T=10^4$ for $N<50$ and $T=10^3$ for $N\ge 50$.

\subsection{Small systems}
\begin{figure}
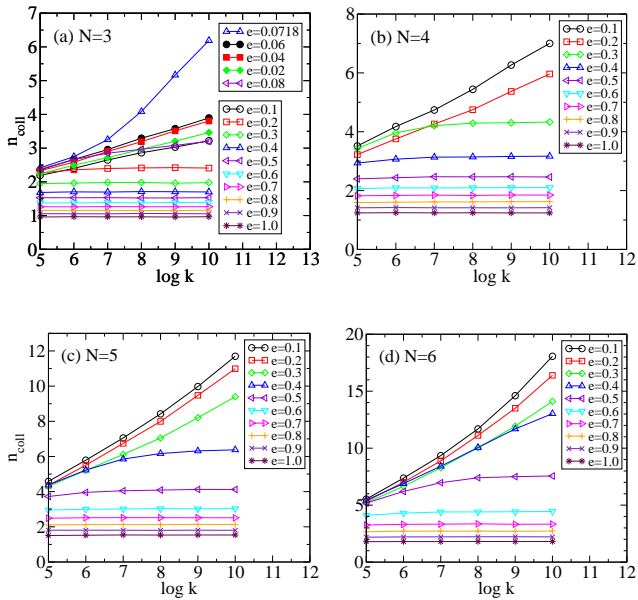

\subfigure{
\includegraphics[width=0.48\hsize,clip]{./collfreq-logk_N3.eps}
\label{cflogk3}}
\subfigure{
\includegraphics[width=0.45\hsize,clip]{./collfreq-logk_N4.eps}
\label{cflogk4}}
\subfigure{
\includegraphics[width=0.48\hsize,clip]{./collfreq-logk_N5.eps}
\label{cflogk5}}
\subfigure{
\includegraphics[width=0.45\hsize,clip]{./collfreq-logk_N6.eps}
\label{cflogk6}} \caption{Collision rate per particle $n_{\rm coll}$ vs.
$\log k$ for various values of the restitution
coefficient $e$ for the systems with $N=3$ (a), 4 (b), 5 (c), and 6 (d). } \label{cflogk}
\end{figure}

Let us first consider systems with small number of particles, in which a
series of collisions occurs in a simple manner.
Figure~\ref{cflogk} shows the
collision rates per particles, $n_{\rm coll}$ as a function of $k$ for
various values of $e$ on the system with $N=3\sim 6$.
For the system of $N=3$ (Fig.~\ref{cflogk3}), the logarithmic
behavior of $n_{coll}$ is clearly observed for $e<e_{c}(3)=7-4\sqrt{3}$,
as has been suggested from the analysis in Sec.~IID.
On the other hand, for $e>e_{c}(3)$, $n_{\rm coll}$ converges to a finite value as $k$ becomes
large.  It should be noted, however, that $n_{\rm coll}$ increases
faster than $\log k$ for $e=0.0718\simeq e_{c}(3)$.

For the systems with $N=$4, 5, and 6, such a region where
$n_{\rm coll}$ increases faster than $\log k$ extends toward the smaller
$e$ region than $e_{c}(N)$, $e<e_{c}(N)$, as is seen in Fig.~\ref{cflogk}.
Here, $e_{c}(N)$ represents the critical
restitution coefficient of the inelastic collapse for the free $N$ particle system
evaluated by the ICW model with Eq.~(\ref{icw}) and
Eq.~(\ref{symmetry}),
\begin{eqnarray}
 e_{c}(N)=\tan^2\left(\frac{\pi}{4}\left(1-\frac{2}{N}\right)\right),
\end{eqnarray}
which we expect accurate for small $N$.

\begin{figure}
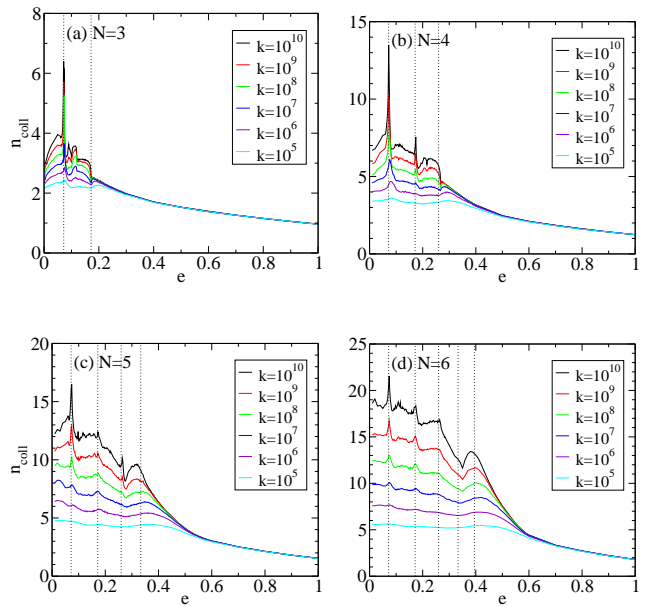

\subfigure{
\includegraphics[width=0.48\hsize,clip]{./collfreq-e_N3.eps}
\label{cfe3}}
\subfigure{
\includegraphics[width=0.45\hsize,clip]{./collfreq-e_N4.eps}
\label{cfe4}}
\\
\subfigure{
\includegraphics[width=0.48\hsize,clip]{./collfreq-e_N5.eps}
\label{cfe5}}
\subfigure{
\includegraphics[width=0.45\hsize,clip]{./collfreq-e_N6.eps}
\label{cfe6}}
\caption{Collision rate per particle $n_{\rm coll}$ vs. $e$, plotted for
$k=10^{5}\sim 10^{10}$ for the systems with
$N=3$ (a), 4 (b), 5 (c), and 6 (d). The vertical dotted lines are
at the restitution coefficient of
$e_{c}(3)$, $e_{c}(4)$, $\cdots$, $e_{c}(N+1)$ from left to right.
\label{cfe}
}
\end{figure}

In Fig.~\ref{cfe}, $n_{\rm coll}$ is shown as a function of $e$ for
$k=10^5 \sim 10^{10}$ on the system with $N=3\sim 6$.
One may notice that the curves have irregular-looking fine
structures.  We confirmed, however, that their statistical errors are
small enough and that these fine structures are reproducible if we
change sequences of random numbers in the simulations.
Some of larger structures coincide with the critical
restitution constants $e_c(n)$ with ($n=3,\cdots ,N+1$),
which are shown by the vertical dotted lines in Fig.~\ref{cfe}.
%
In the case of $N=3$ (Fig.~\ref{cfe3}),
one can observe a sharp peak at $e_{c}(3)$, 
and the peak value of
$n_{\rm coll}$
increases faster than $\log k$ as $k$ increases.  For $e<e_{c}(3)$,
$n_{\rm coll}$ increases by a nearly constant when $k$ becomes 10 times
larger, which suggests that $n_{\rm coll}$ increases logarithmically as
is discussed above.

For $N=3 \sim 6$, there are a couple of features in common.
First, a sharp peak appears at $e_{c}(n)
(n=3,\cdots ,N-1)$ and the peak at $e_{c}(3)$ is highest.
Secondly, a dip appears at $e$ slightly larger than $e_{c}(N)$. Our
simulation results (not shown here) suggest that this dip still exists
for $N=10$, becomes unclear for $N=25$, and
completely disappears for $N=30$.

\subsection{Large systems}

\subsubsection{Collision rate in the large $k$ limit}

\begin{figure}
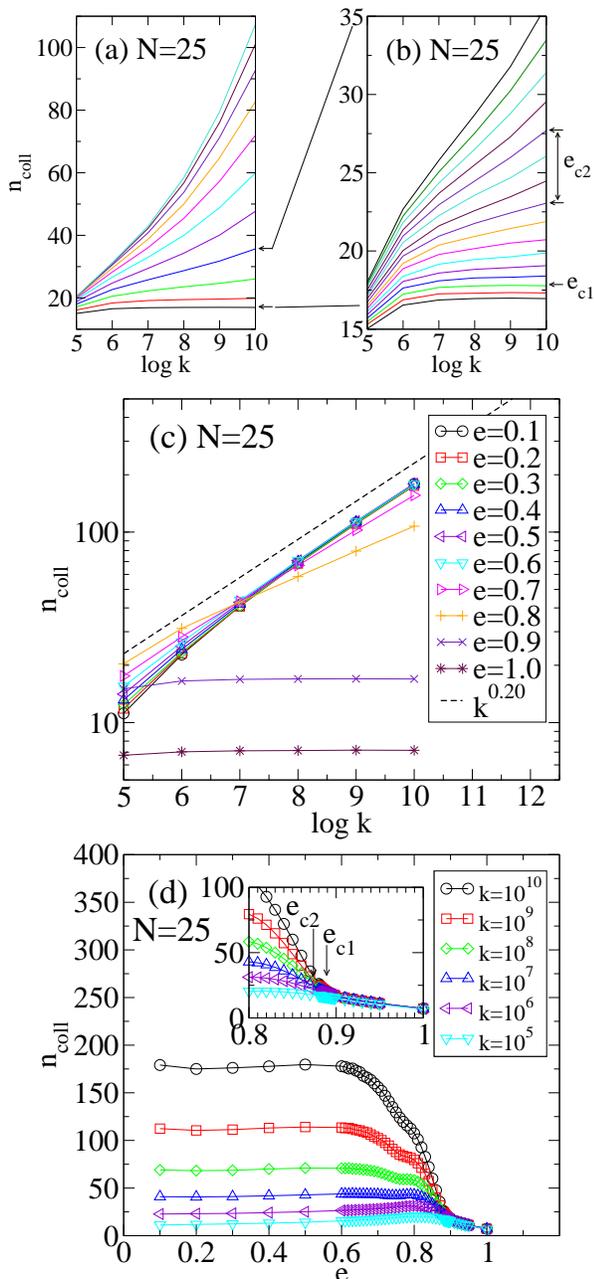

\includegraphics[width=0.9\hsize,clip]{./cfreq-log_N25.eps}\\[0.2cm]
\includegraphics[width=0.8\hsize,clip]{./collfreq-logk_N25.eps}
\includegraphics[width=0.8\hsize,clip]{./collfreq-e_N25.eps}
\caption{
Collision rate per particle $n_{\rm coll}$ for $N=25$.
$n_{\rm coll}-\log k$ is plotted for
(a) $e=0.80 \sim 0.90$ with an increment 0.01 from top to bottom.
(b) $e=0.87 \sim 0.90$ with an increment 0.002 from top to bottom.
$\log n_{\rm coll}-\log k$ is plotted for
(c) $e=0.1 \sim 1.0$ with an increment 0.1.
(d) $n_{\rm coll}-e$ is plotted for $k=10^5 \sim 10^{10}$ from bottom to
top with the inset that shows in close-up near the critical values
$e_{c1}\approx 0.894$ and $e_{c2}\approx 0.88$.}  \label{cfreqlogN25}
\end{figure}
For large systems, the power-law divergence of the collision rate
dominates, but we can see clearly that there
exists the region of restitution coefficient where $n_{\rm coll}$
diverges definitely slower than the power law.

In Fig.~\ref{cfreqlogN25}, we plot $n_{\rm coll}$ for $N=25$ as a
function of $k$ for various values of $e$ (a, b, and c), and as a
function of $e$ for various values of $k$ (d).
It is clear in the logarithmic plot of
Fig.~\ref{cfreqlogN25}(c) that $n_{\rm coll}$ converges for $e\gtrsim
0.9$ and diverges in the power-law for $e\lesssim 0.8$.
The exponent $\alpha$ in the power-law regime depends on $e$,
but is nearly constant $\alpha\simeq 0.2$ for $e\lesssim 0.6$ as can be
seen from Fig.~\ref{cfreqlogN25}(c).
In Fig.~\ref{cfreqlogN25}(d), we also observed that the value of
$n_{\rm coll}$ itself is nearly independent of $e$ for
any value of $k$ for $e \lesssim 0.6$, where the exponent $\alpha$ is
nearly constant.

\begin{figure}
\begin{center}
\includegraphics[width=0.8\hsize,clip]{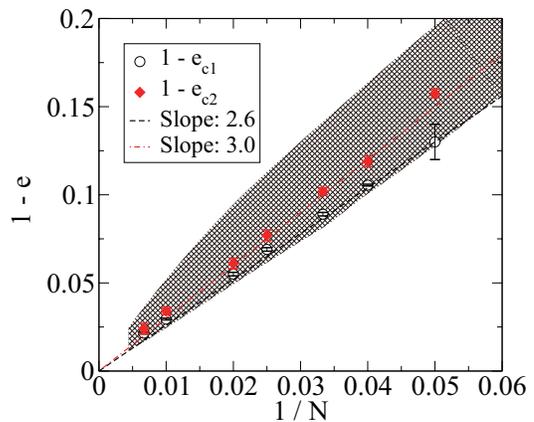}
\caption{
Bifurcation diagram for the three regimes.
The critical values of $e_{c1}$ and $e_{c2}$ are plotted for 
 $N=$20, 25, 30, 40, 50, 100, and 150.
The error bars show ambiguity of the results in the procedure.
The dashed and the dashed dotted lines show the fitting lines
for $e_{c1}$ and $e_{c2}$ by
$(1-e_{c1}^{\rm fit})={2.6/ N}$ and
$(1-e_{c2}^{\rm fit})={3.0/ N}$, respectively.
The shaded region represents the region where
 the partially condensed state appears; its boundary is estimated at
$N=$20, 25, 30, 50, 100 and 150.
}
\label{bifurcationdiagram}
\end{center}
\end{figure}

In the following, we will examine the transition region between
the converging regime to the diverging regime carefully.  Let us denote
the lower limit of the restitution of the converging region by $e_{c1}$
and the upper limit of the power-law diverging region by $e_{c2}$.
A close look at the region $0.8 < e < 0.9$ in the semi-logarithmic plots of
Fig.~\ref{cfreqlogN25}(a, b) reveals that there are two regimes within
the region where $n_{\rm coll}$ diverges: the convex regime and the
concave regime as a function of $\log k$.  In the convex regime, $n_{\rm
coll}$ diverges faster than $\log k$, suggesting that it is a part of
the power-law regime.  In the concave regime, the divergence is slower
and it seems that $n_{\rm coll}$ eventually shows the logarithmic
divergence
\begin{equation}
n_{\rm coll}\sim b(e,N)\log k+ \mbox{const.}
\label{b(e,N)}
\end{equation}
in the large $k$ limit with the coefficient $b$ that depends on $e$ and
also on $N$.
%
The lower limit of the converging regime $e_{c1}$ is determined as the
upper limit of the divergence; the data are fitted to Eq.(\ref{b(e,N)})
in the asymptotic region, then $e_{c1}$ is the point where $b=0$.  On
the other hand, the value of $e_{c2}$ is estimated by the boundary
between the convex and the concave regime.  By these procedures, we
obtain that $e_{c1}\simeq 0.894$ and $e_{c2}$ is somewhere between 0.878
and 0.884 for $N=25$.
The values of $e_{c1}$ and $e_{c2}$ are determined for several values of
$N$, and plotted with error bars in the $1/N$-$(1-e)$ plane in
Fig.~\ref{bifurcationdiagram}.  One can see that they fit very well to
the lines
\begin{equation}
(1-e_{c1}^{\rm fit})={2.6\over N}, \quad
(1-e_{c2}^{\rm fit})={3.0\over N}.
\label{e-fit}
\end{equation}
Note that their functional form is the same with the asymptotic form of
$e_{c,{\rm ICW}}^{\rm wall}$ in Eq.~(\ref{ec_icw_wall}).

\begin{figure}
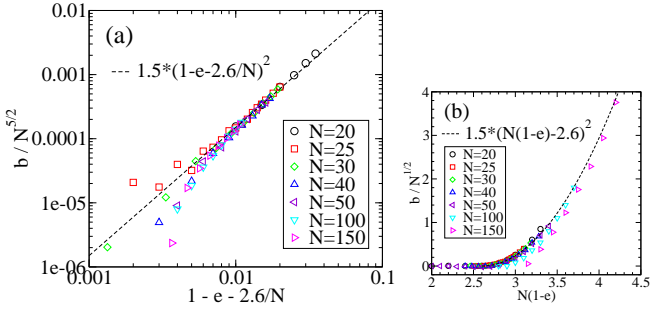

\begin{center}
\includegraphics[width=0.6\hsize,clip]{./bsc_Xfit_rev.eps}
\includegraphics[width=0.38\hsize,clip]{./logfit_b-X.eps}
\end{center}
\caption{(a) $b(e,N)/N^{5/2}$ vs. $(1-e)-2.6/N$ in the logarithmic
scale.  $b(e,N)$ is defined in Eq.(\ref{b(e,N)}) and estimated from the
plots similar to those in Fig.\ref{cfreqlogN25}(c) for various values of
$N$ and $e$ in the logarithmically collapsing regime.  The dashed line
gives a fit by $1.5((1-e)-2.6/N)^2$ for $b/N^{5/2}$.
(b) $b(e,N)/N^{1/2}$ vs. $N(1-e)$ in the linear scale using the same data
as in (a).
The range of values of $e$ used to plot (a) and (b) is the following:
$0.835\le e \le 0.900$ for $N=20$,
$0.876\le e \le 0.900$ for $N=25$,
$0.894\le e \le 0.920$ for $N=30$,
$0.918\le e \le 0.940$ for $N=40$,
$0.932\le e \le 0.960$ for $N=50$,
$0.963\le e \le 0.972$ for $N=100$,
and
$0.972\le e \le 0.979$ for $N=150$.
\label{logfitbX}}
\end{figure}

We further examine $b(e,N)$ in Eq.(\ref{b(e,N)}) as a function of
both $e$ and $N$ by simulation data. From Eq.(\ref{e-fit}), we expect
that $b(e,N)$ is expressed by a simple function of $(1-e)-A/N$ with a
constant $A\approx 2.6$.
This is actually what we find in Fig.~\ref{logfitbX}(a), where we plot
$b(e,N)/N^{5/2}$ against $(1-e)-2.6/N$ for various values of $N$ and $e$
in the logarithmic scale.  One can see that the data collapse on a
straight line with the slope 2, which means
\begin{equation}
b(e,N) \sim N^{5/2}\Bigl((1-e)-2.6/N\Bigr)^2,
\label{b_e_N}
\end{equation}
from which we confirm the asymptotic form
\begin{equation}
e_{c1} \simeq  1-{2.6\over N}.
\label{e_c1-asympt}
\end{equation}
It should be noted that
this result is very close to the critical values of $e$ below which clustering starts in the 1-d
granular system driven by a vibrating bottom plate, i.e.
$e_c\approx 1- 2.5/N$
\cite{LudingClementBlumenRajchenbachDuran-1994} and
$e_c\approx 1- 2.6/N$ \cite{BernuDelyonMazighi-1994}.

Based upon above analysis, we conclude that the critical values,
$e_{c1}$ and $e_{c2}$, do not coincide; therefore, in addition to the
converging regime, 
there are two diverging regimes in
$e$ with respect to behavior of $n_{\rm coll}$ in the hard-sphere limit
$k\to\infty$:
(i) {\it uncollapsing regime}, $e > e_{c1}$, where $n_{\rm coll}$
converges to a constant value, (ii) {\it logarithmically collapsing
regime}, $e_{c1}> e > e_{c2}$, where $n_{\rm coll}$ diverges as $n_{\rm
coll}\sim\log k$, and (iii) {\it power-law collapsing regime},
$e<e_{c2}$, where $n_{\rm coll}$ diverges as $n_{\rm coll}\sim
k^\alpha$.  

\subsubsection{Partially condensed state}

\begin{figure}
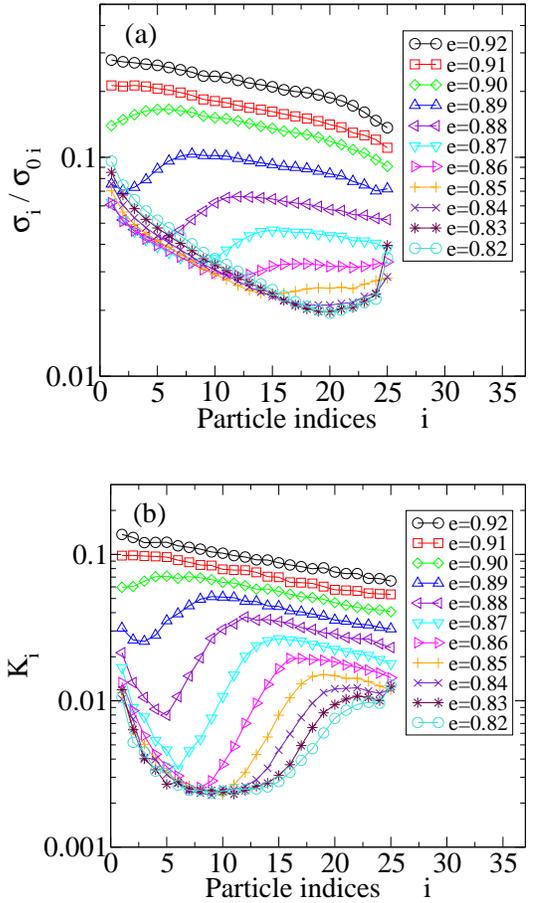

\subfigure{
\includegraphics[width=0.80\hsize,clip]{./scal-sdev_fine.eps}
\label{scalsdev}}
\\
\subfigure{
\includegraphics[width=0.80\hsize,clip]{./kinetic_fine.eps}
\label{kinetic}}
\caption{ Spatial variation of the particle fluctuation of position
and the kinetic energy for $e=0.82\sim 0.92$ for the system with $N=25$.  (a) The standard
deviation of particle position $\sigma_i$ of particle $i$, normalized by
the corresponding value for $e=1$,  $\sigma_{0,i}$.  (b) The
kinetic energy $K_i$ of the particle $i$.  The data are shown for the
cases with the elastic constant $k=10^{10}$.  \label{partial_cond}}
\end{figure}

We observe a partial condensation near
the bottom around a certain value of $e$.
Figure~\ref{partial_cond} shows the spatial variation of the positional
fluctuation (a) and the kinetic energy (b) of each particle for various
values of $e$ for $N=25$; Fig.~\ref{scalsdev} shows the standard
deviation $\sigma_i$ of the position of the particle $i$ divided by
the one $\sigma_{0\,i}$ for the elastic case $e=1$, and (b)
shows the kinetic energy $K_i$ of the particle $i$.  Note that
$K_i=(1/2)k_{\rm B} T_0$ for any particle when $e=1$.

For $e\gtrsim 0.91$, $\sigma_{i}/\sigma_{i\,0}$ and
$K_i$ are larger in the region closer to the bottom and they decrease
monotonically as the particle index $i$ increases. This is
because the thermal wall at the bottom supplies the kinetic energy to
the bottom particle, and the kinetic energy is
dissipated as it is transported away from the bottom via the
inelastic collisions.
However, around $e\simeq 0.9$, a dip appears near the bottom both in
$\sigma_{i}/\sigma_{0\,i}$ and $K_i$, and there appears the inversion
layer where the temperature increases with $i$. This means that the low
temperature and high density domain appears near the bottom.  We call
this {\it the partially condensed state}.  For $N=25$, the value of
$e\simeq 0.9$ where the partially condensed state appears almost
coincides with, but seems to be slightly larger than 
$e_{c1}\simeq 0.894$, i.e. the critical point of the inelastic collapse.

The condensed domain with low $\sigma_{i}/\sigma_{i\,0}$ in
Fig.~\ref{scalsdev} extends towards the upper part of the system as $e$
is decreased down to $0.84$, where the whole system is condensed.  That
is, the partially condensed state appears for $0.84\lesssim e\lesssim
0.90$, namely, the partially condensed state appears both in the
logarithmically collapsing and the power-law collapsing regimes.

In Figure~\ref{bifurcationdiagram}, the region for the partially
condensed state is shown by the shaded area in the $1/N$-$(1-e)$ plane.
One can see that the upper bound of $e$ (the lower boundary in
Fig.~\ref{bifurcationdiagram}) for the partially condensed state nearly
coincides with $e_{c1}$ for all of the cases examined.  

\subsubsection{Exponent in the power-law collapsing regime}

\begin{figure}
\begin{center}
\includegraphics[width=0.75\hsize,clip]{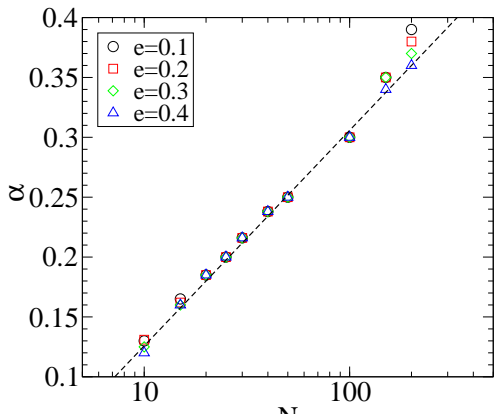}
\caption{Exponent of the power law $\alpha$ for $e=0.1,\,0.2,\,0.3,$
and $0.4$ plotted as a function of $N$.  The dashed line gives a fit by
$0.18 \log N -0.054$.}  \label{powerexpo}
\end{center}
\end{figure}

We observed that $n_{\rm coll}$ is almost constant when $e\le 0.4$ for
$N=25$ as can be seen in Fig~\ref{cfreqlogN25}.  We examined this for
the case with $N=20\sim 150$, and found this is true for all the cases.
In Fig.~\ref{powerexpo}, the exponents $\alpha$ for $e=$0.1, 0.2, 0.3
and 0.4 are plotted against $\log N$.  One can see that $\alpha$ does
not change by $e$ but depends linearly on $\log N$ and can be fitted to
\begin{eqnarray}
\alpha^{\rm fit}= 0.18\log N-0.054 \simeq 0.18\log\left(N/2\right).
\label{alphafit}
\end{eqnarray}
This logarithmic dependence of
the exponent $\alpha$ on $N$
means that $n_{\rm coll}$ is given by
\begin{eqnarray}
n_{\rm coll} \sim (N/2)^{0.18 \log k},
\end{eqnarray}
for $e\le 0.4$.

\subsubsection{Effect of the floor temperature $T_0$}
We find no qualitative difference
in the $k$-dependence of $n_{\rm coll}$ by changing $k_{\rm B} T_0$
from 1 to 10
in both systems with $N=25$ and $50$.
We show in Table~\ref{table1} the critical values $e_{c1}$ and $e_{c2}$, and the average exponent of the power law
$\overline{\alpha}$ for $k_{\rm B} T_0=1$ (see Sec.~IIIC.1),
$2$, $4$, and $10$ for the systems with $N=25$ and $50$.
Here $\overline{\alpha}$ is the arithmetic mean of the four values of $\alpha$ at $e=0.1,\, 0.2,\, 0.3,$ and $0.4$,
for which $\alpha$'s are almost constant independent of $e$.
The both critical values and the power law exponent seem to be independent of $T_0$.
\begin{table}[htbp]
\catcode`?=\active \def?{\phantom{0}}
\begin{tabular}{ccccc}\hline
??$N$ & ??$k_{\rm B} T_0$??& ??$e_{c1}$?? & ??$e_{c2}$?? & ???????$\overline{\alpha}$????? \\ \hline
??25 & $1$ & $0.894$ &??$0.878 \sim 0.884$ &??0.20 \\
?? & $2$ & $0.892$ &??$0.874 \sim 0.880$ & ??0.20  \\
?? & $4$ & $0.892$ &??$0.872 \sim 0.880$ & ??0.20  \\
?? & $10$ & $0.890$ &??$0.872 \sim 0.878$ & ??0.19  \\
??50 & $1$ & $0.944$ &??$0.942 \sim 0.936$ &??0.25\\
?? & $2$ & $0.944$ &??$0.938 \sim 0.942$ &??0.25 \\
?? & $4$ & $0.944$ &??$0.938 \sim 0.940$ &??0.24  \\
?? & $10$ & $0.944$ &??$0.934 \sim 0.944$ &??0.24  \\ \hline
\\
\end{tabular}
\caption{\label{table1}
The critical values $e_{c1}$, $e_{c2}$ and the average exponent of the power law $\overline{\alpha}$ for various values of the floor temperature $T_0$ for the systems with $N=25$ and $50$. The average exponent $\overline{\alpha}$ is defined as the arithmetic mean of the four values of $\alpha$ at $e=0.1,\, 0.2,\, 0.3,$ and $0.4$.}
\end{table}

\subsection{Inelastic collapse in the free space}

In order to narrow down the possible origin of the power law divergence
of the collision rate, we further simplify the system and consider the
$N$-particle system in the 1-d free space without the external drive and
the gravity.  We performed MD simulations to see how the total number of
collisions behaves in the hard sphere limit.

In the initial state, $N$ particles of the diameter $d$ are placed at an
equal interval with the space $a$,
\begin{equation}
x_i = \left(i-{N+1\over 2}\right)(d+a);
\quad i=1, 2, \cdots N,
\end{equation}
with the initial velocities,
\begin{equation}
v_i = -v_0\, {\rm sgn}(x_i) + \delta v_0\, \xi_i ,
\end{equation}
where ${\rm sgn}(x)$ is the sign function and $\xi_i$'s are random
numbers distributed uniformly over the interval $[-1,1)$;  $v_0$ and
$\delta v_0$ are positive parameters.

\begin{figure}
\subfigure{
\includegraphics[width=0.8\hsize,clip]{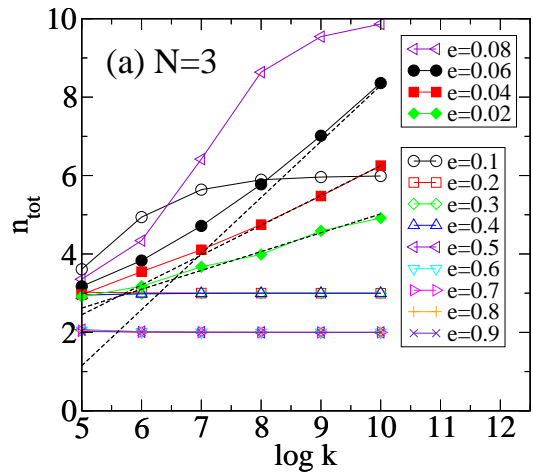}
\label{ntot3}}
\subfigure{
\includegraphics[width=0.82\hsize,clip]{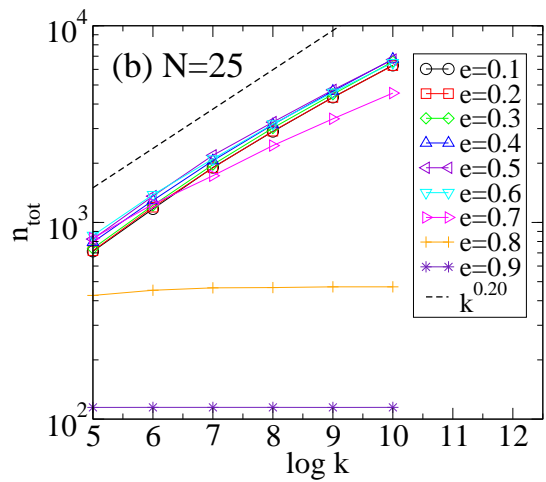}
\label{ntot25}}
\caption{$n_{\rm tot}$ vs. $\log k$ for various values of $e$ (a) for
$N=3$ in the log-linear scale and (b) for $N=25$ in the log-log scale.
The parameters for the initial state are $a=v_0=1$ and $\delta v_0=0.1$.
Each data point represents an average over $1000$ realizations by
different random number $\xi_i$'s.  The dashed lines in (a) show the
asymptotic behavior given by Eq.(\ref{ntot-log}) with $q$ given by
Eq.(\ref{def_q}) for the corresponding $e$ values after the constants being adjusted to the data.  The
dashed line in (b) is the line with the slope 0.2.  } \label{ntot}
\end{figure}

We count the total number of collisions $n_{\rm tot}$ until the
relative velocity of the end particles $v_N-v_1$ becomes positive.
The results are shown in Fig.\ref{ntot} for the systems with $N=3$ (a)
and 25 (b) for $d=a=v_0=1$ and $\delta v_0=0.1$.  One can see that the
total number of collisions $n_{\rm tot}$ behaves in an analogous way with
the collision rate $n_{\rm coll}$ in the driven system under the gravity
shown in Figs.\ref{cflogk3} and \ref{cfreqlogN25}(c).
For the case of $N=3$,  $n_{\rm tot}$ converges to a finite value when
$e>e_c(3)\simeq 0.0718$ and diverges as $\log k$ when $e<e_c(3)$.
The dashed lines in Fig.\ref{ntot3} show Eq.(\ref{ntot-log}) with
$q$ given by Eq.(\ref{def_q}) for the corresponding $e$ values, and adjusted constants.
As for the case of $N=25$, $n_{\rm tot}$ diverges as $k^\alpha$ when
$e\lesssim 0.6$ with the exponent close to the value $\alpha\simeq 0.2$
for the previous case with the drive and the gravity.

\section{Summary and Discussions}


We have studied the inelastic collapse in the 1-d system under the
gravity with the random driving from the bottom floor.  Using MD
simulations for the soft-sphere systems, we calculated the collision
rate per particle $n_{\rm coll}$ and see how it diverges/converges in
the hard-sphere limit; The hard-sphere limit is taken by the infinite
limit of the elastic constant, $k\to\infty$ with the restitution
coefficient $e$ being kept constant.  We have found that there are three
regimes in the restitution coefficient $e$: (i) {\it the uncollapsing
regime} for $1\ge e>e_{c1}$, where $n_{\rm coll}$ converges, (ii) {\it
the logarithmically collapsing regime} for $e_{c1}> e > e_{c2}$, where
$n_{\rm coll}$ diverges as $n_{\rm coll}\sim \log k$, and (iii) {\it the
power-law collapsing regime} for $e_{c2}>e>0$, where $n_{\rm coll}\sim
k^\alpha$.
For small $N$ systems, the region of $e$
for the power-law collapsing regime is small and disappears for $N=3$.
On the other hand, for large $N$ systems, the region for the power-law
collapsing regime expands in the way that both of $e_{c1}$ and $e_{c2}$
approaches 1 as Eq.(\ref{e-fit}).
As for the floor temperature effect, we have checked the critical restitution
constants $e_{c1}$ and $e_{c2}$ and the power law exponent $\alpha$
for the system of $N=25$ and $50$, and found virtually no change for all of them
in the temperature range of $1\le k_{\rm B} T_0\le 10$.

If the intervals of collisions follow a geometrical sequence toward the
inelastic collapse, the logarithmic divergence of the {\it collision
number} can be understood based on the consideration that the collision sequence
terminates at the point where the collision interval becomes comparable with the
duration time of binary collision.  In the case of one particle bouncing
on a floor under the gravity, it is obvious that the collision times
follow the geometrical sequence.  We have shown that it holds also for
the three-particle system in the 1-d free space.  Thus, the logarithmic
divergence of the {\it collision rate} for the externally driven system
can be understood if the inelastic collapses occur at a certain rate and
they do not interfere each other nor are affected by the external drive
in the hard-sphere limit, $k\to\infty$.

On the other hand, the power-law divergence of the collision rate is
more intriguing.  It has been reported in the gravitational slope
flows~\cite{MitaraiNakanishi-2003}, but our results show that it occurs
in an even simpler system, i.e. a 1-d externally driven system under the
gravity.  If we try to understand this in the same way as above, the
collision interval should decrease as a power of collision number.  This
possibility is supported by the fact that the total number of collisions
in the free space also diverges in the power law in the hard sphere
limit.  If this is true, we still need to understand how the power law
sequence collision intervals arises.
%

For the small $N$ systems, the collision rate shows a certain structure
as a function of $e$ at $e=e_c(n)$ of $3\le n\le N+1$, i.e. the critical
restitution coefficient for the $n$-particle system in the free space.
For $N=3$, there is a sharp peak at $e=e_c(3)$ and a dip at $e=e_c(4)$,
while for $3<N\le 6$, we find peaks at $e=e_c(n)$ for $3\le n\le
N-1$, a dip or a shoulder at $e\simeq e_c(N)$, and a somewhat broad
peak around $e\simeq e_c(N+1)$.  Such a structure becomes vague for
larger $N$.
Note that the relative motion of $n$
particles under gravity with respect to their center of mass is
equivalent to the motion in the free space as long as they do not
interact with their surrounding particles.  Thus, the structure at
$e_c(n)$ for $n<N$ should be an effect of the inelastic collapse in
which a part of the system is involved, but we do not understand yet how
it shows up as a sharp peak or a dip structure, depending on the number
of particles involved.

For large $N$, it is also intriguing that the collision rate is
independent of $e$ for a rather wide range; In the case of $N=25$,
$n_{\rm coll}$ is constant in the region $e\lesssim 0.6$
for any $k$, thus the exponent $\alpha$ also does not depend on $e$ in the same region.
Within the range $20\le N \le 150$,
$n_{\rm coll}$ is almost constant for $e\le 0.4$, but the power-law exponent
increases depending linearly on $\log N$ as $N$ is increased.

One may find that some data points for $N>100$ in
Figs.~\ref{bifurcationdiagram}, \ref{logfitbX}(b), and \ref{powerexpo}
seem to deviate systematically from the asymptotic fitting expressions
given by Eqs.~(\ref{e-fit}), (\ref{b_e_N}), and (\ref{alphafit}),
respectively.  This may be due to the  excessive load on particles near
the floor.
In the case of very large $N$, the load on the particles near the floor
becomes so large that the contacts among them cannot be resolved into
binary collisions but may remain as a long-lived contacts in the range
of $k$ investigated. In such a case, the system behavior
may deviate from the assumed asymptotic forms.


It is found that there appears the partially condensed state, where
some particles near the bottom condense with lower kinetic energy.
The region where the partially condensed
state appears in the $N$-$(1-e)$ plane covers the logarithmically collapsing regime; It starts
hardly inside the uncollapsing regime and extends somewhat into the
power-law collapsing regime.  The condensed state near $e_{c1}$
contains only a few particles, but the number of condensed
particles increases as $e$ decreases.  The fact that the partially
condensed state starts almost at $e_{c1}$ suggests that the
inelastic collapse causes the condensation, but it remains to be
understood how a small number of particles can condense by the
inelastic collapse at $e\approx e_{c1}$, that is much larger than
$e_c(n)$ for small $n$.

The partially condensed state has already been observed in the 1-d
granular systems driven by a vibrating bottom plate in various forms of
vibration. In the case of the sinusoidal vibration, the condensed state
has been shown to appear for $N(1-e)\agt
2.5$~\cite{LudingClementBlumenRajchenbachDuran-1994}, and in the cases
of a sawtooth vibration and a piecewise quadratic vibration, for
$N(1-e)\agt 2.6$~\cite{BernuDelyonMazighi-1994}.  Our result shows that
the partially condensed state appears below $e_{c1}$, which means
$N(1-e)\agt 2.6$ from Eq.~(\ref{e_c1-asympt}). These results are
consistent with each other and show that the point where the system
starts to condense is not sensitive to the driving mode.

In summary, we have demonstrated that the inelastic collapse shows up in
the 1-d driven system under the gravity as the diverging collision rate
in the large $k$ limit with keeping the restitution coefficient $e$
constant.  By numerical simulations, we found that there are three
regimes for the way that the collision rate diverges, i.e.  the
uncollapsing regime, the logarithmically collapsing regime, and the
power-law collapsing regime.

\appendix
\section{}

In this appendix, we consider the three-body inelastic collapse of
the inelastic hard spheres in the free space and derive the
asymptotic behavior Eq.~(\ref{deltat12}) of the time $t_{12}^{(n)}$ between two
successive collisions between the particles 1 and 2.

\begin{figure}[htbp]
\includegraphics[width=0.9\hsize,clip]{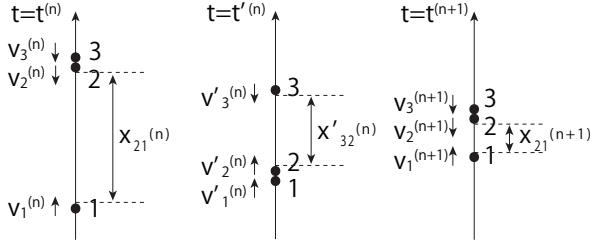}

\vspace{0.5cm}
\caption{\label{collisions}
Schematic picture of collision sequence occurring in the
three-body inelastic collapse in the free space.}
\end{figure}

Let $t^{(n)}$ be the time of the $n$th collision
between the particles 2 and 3 and $t'^{(n)}$ be the
time of the $n$th collision between the particles 1
and 2, and define $t_1^{(n)}={t'}^{(n)}-t^{(n)}$ and
${t}_2^{(n)}=t^{(n+1)}-t'^{(n)}$ (See, Fig.~\ref{collisions}).
Similarly, the particle velocities just after $t^{(n)}$ are
$v_i^{(n)}\,(i=1,2,3)$, those just after ${t'}^{(n)}$ are
${v'}_i^{(n)}\,(i=1,2,3)$.  The relative velocities are
denoted by $v_{21}^{(n)}=v_2^{(n)}-v_1^{(n)}$ and
$v_{32}^{(n)}=v_3^{(n)}-v_2^{(n)}$.
The separation between the particles 1 and 2 at $t^{(n)}$ is
denoted by $x_{21}^{(n)}$ and that between the particles 2 and 3 at
${t'}^{(n)}$ by ${x'}_{32}^{(n)}$.

\vspace{0.5cm}

The time $t_1^{(n)}$ and $t_2^{(n)}$ are then written as
\begin{eqnarray}
 t^{(n)}_1&=&\frac{x^{(n)}_{21}}{|v^{(n)}_{21}|},
\hspace{1cm}
 {t}^{(n)}_2=\frac{{x'}^{(n)}_{32}}{|{v'}^{(n)}_{32}|},
\label{t1t2}
\end{eqnarray}
respectively. The separations $x_{21}^{(n)}$ and ${x'}_{32}^{(n)}$ can
also be expressed as
\begin{eqnarray}
x^{(n)}_{21}={v'}^{(n-1)}_{21} {t}^{(n-1)}_{2},
\hspace{1cm}
{x'}^{(n)}_{32}=v^{(n)}_{32} t^{(n)}_{1},
\label{x21x32}
\end{eqnarray}
respectively.  Combining Eqs.~(\ref{t1t2}) and (\ref{x21x32}), we can
write
\begin{eqnarray}
 t^{(n)}_1&=&\frac{{v'}^{(n-1)}_{21}}{|v^{(n)}_{21}|}\cdot\frac{v^{(n-1)}_{32}}{|{v'}^{(n-1)}_{32}|}\, t^{(n-1)}_1
\end{eqnarray}
Using $v^{(n-1)}_{21}=-\frac{1}{e}{v'}^{(n-1)}_{21}$ and $v^{(n)}_{32}=-e{v'}^{(n-1)}_{32}$, we can further rewrite it as
\begin{eqnarray}
 t^{(n)}_1&=&\frac{1}{e^2}\frac{{v}^{(n)}_{32}}{|v^{(n)}_{21}|}
\cdot\frac{{v}^{(n-1)}_{32}}{|v^{(n-1)}_{21}|}
\left(\frac{v'^{(n-1)}_{21}}{|{v'}^{(n-1)}_{32}|}\right)^2
 t^{(n-1)}_1.
\end{eqnarray}
This relation can be expressed using the ratio of the relative
velocities $m^{(n)}=v^{(n)}_{32}/v^{(n)}_{21}$ and
${m'}^{(n)}={v'}^{(n)}_{32}/{v'}^{(n)}_{21}$ as
\begin{eqnarray}
 t^{(n)}_1&=&\frac{1}{e^2}\frac{m^{(n)} m^{(n-1)}}{({m'}^{(n-1)})^2}\,
 t^{(n-1)}_1.
\label{t1mm}
\end{eqnarray}

The sequence of $m^{(n)}$ and ${m'}^{(n)}$, which are completely
determined by the collision laws and the initial condition $m^{(0)}$ (or ${m'}^{(n)}$),
has been studied by Constantin et
al.~\cite{ConstantinGrossmanMungan-1995}.  We briefly summarize their
results that are relevant for our purpose in this appendix.  Upon a
collision between particles 1 and 2, velocities after the collision
$(v'_1,v'_2)$ and before the collision $(v_1,v_2)$ are related by
\begin{eqnarray}
 \left(
\begin{array}{c}
 v'_1\\
 v'_2
\end{array}
\right)
&=&
 \left(
\begin{array}{cc}
 \frac{1-e}{2}&\frac{1+e}{2}\\
 \frac{1+e}{2}&\frac{1-e}{2}
\end{array}
\right)
 \left(
\begin{array}{c}
 v_1\\
 v_2
\end{array}
\right).
\end{eqnarray}
From this collision law, we can deduce the following relations:
\begin{eqnarray}
 {m}^{(n)}&=&-e\frac{{m'}^{(n-1)}}{1+b\,{m'}^{(n-1)}},
\label{mmdash}
\\
 {m'}^{(n)}&=&-\frac{1}{e}\left(m^{(n)}+b\right),
\label{mdashm}
\end{eqnarray}
where $b\equiv(1+e)/2$.  If $e<e_c(3)=7-4\sqrt{3}$ and $m^{(0)}$ (or
${m'}^{(0)}$) is such that the collision sequence continues
infinitely, i.e. the inelastic collapse occurs, then
$m^{(n)}$ and $m'^{(n)}$ should converge to
the stable fixed point values
\begin{eqnarray}
 m^{*}&=&\frac{1}{2}\left(-b+\sqrt{b^2-4e}\right),
\\
 {m'}^{*}&=&\frac{1}{2e}\left(-b-\sqrt{b^2-4e}\right),
\end{eqnarray}
which are real when $e\le e_c(3)$.

Therefore, if both $n$ and $n'$ are so large that $m^{(n)}\approx
m^{(n')}\approx m^{*}$ and ${m'}^{(n)}\approx
{m'}^{(n')}\approx{m'}^{*}$, Eq.~(\ref{t1mm}) can be written as
\begin{eqnarray}
 t^{(n)}_1
\approx
\left(\frac{m^{*}}{e\,{m'}^{*}}\right)^2
 t^{(n-1)}_1
\approx
\left(\frac{m^{*}}{e\,{m'}^{*}}\right)^{2(n-n')}
 t^{(n')}_1.
\end{eqnarray}
It is straightforward to show
\begin{equation}
\left(\frac{m^{*}}{e\,{m'}^{*}}\right)^2
=
   {1-6e+e^2 - (1+e)\sqrt{1-14e+e^2} \over
               1-6e+e^2 + (1+e)\sqrt{1-14e+e^2}}
\equiv q,
\label{def_q}
\end{equation}
and
$0\le q <1$ for $0\le e < e_{c}(3)$.
%
%
If we start to count the number of
collisions at $n'$, namely we put $n'=0$,
then we have
\begin{eqnarray}
 t^{(n)}_1 \approx q^n\,  t^{(0)}_1.
\end{eqnarray}

Because of symmetry with regard to exchange of particles,
${t}^{(n)}_2$
should have the same property, and we finally obtain
\begin{eqnarray}
 t_{12}^{(n)}={t}^{(n)}_1+{t}^{(n-1)}_2\approx q^n \, t_{12}^{(0)},
\end{eqnarray}
which is Eq.~(\ref{deltat12}).

\section{}
In this appendix, we turn to the problem of
the three-body inelastic collapse like collisions among the inelastic {\it soft} spheres
and derive the
asymptotic behavior Eqs.~(\ref{t12ss}) and (\ref{deltat12ss}) of the time $t_{12}^{{\rm S}(n)}$
between the instants of the end of contact at two successive collisions between the soft particles 1 and 2.
We put superscript ${\rm S}$ for quantities that are defined for soft particles in this appendix,
in order to distinguish them from the corresponding quantities defined for hard particles in Appendix A.

\begin{figure}
\includegraphics[width=1.0\hsize,clip]{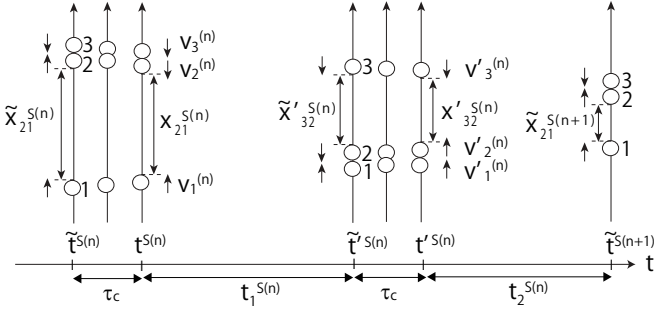}

\vspace{0.5cm}
\caption{\label{collisions-soft}
Schematic picture of collision sequence occurring in the
three-body inelastic collapse like collisions among the inelastic soft spheres in the free space.}
\end{figure}

Let $\tilde{t}^{{\rm S}(n)}$ and $t^{{\rm S}(n)}$ be the times of the beginning and the end of contact at the $n$th collision between the particles 2 and 3, respectively (See, Fig.~\ref{collisions-soft}).
Similarly, let $\tilde{t'}^{{\rm S}(n)}$ and ${t'}^{{\rm S}(n)}$ be the times of the beginning and the end
of contact at the $n$th collision between the particles 1 and 2.
We define the time intervals during which the particles move freely as $t_1^{{\rm S}(n)}\equiv \tilde{t'}^{{\rm S}(n)}-t^{{\rm S}(n)}$
and $t_2^{{\rm S}(n)}\equiv \tilde{t}^{{\rm S}(n+1)}-{t'}^{{\rm S}(n)}$.
Using the duration time $\tau_c$ of contact for a binary collision (see Eq.~(\ref{tauc})),
$t_{12}^{{\rm S}(n)}$ can be represented as
\begin{eqnarray}
 t_{12}^{{\rm S}(n)} = t_1^{{\rm S}(n)}+t_2^{{\rm S}(n-1)}+2\tau_c,
\label{t12s}
\end{eqnarray}
because two collisions occur during the time $t_{12}^{{\rm S}(n)}$.

\vspace{0.5cm}

We denote the relative distance between the particles 1 and 2 just
before the $n$th collision between the particles 2 and 3 as
$\tilde{x}_{21}^{{\rm S}(n)}$, and that just after the collision as
$x_{21}^{{\rm S}(n)}$.  Similarly, the relative distance between the
particles 2 and 3 just before the $n$th collision between the particles
1 and 2 is $\tilde{x'}_{32}^{{\rm S}(n)}$, and that just after the
collision is ${x'}_{32}^{{\rm S}(n)}$. The relative distances just
before and just after a collision are related as follows:
\begin{eqnarray}
 x_{21}^{{\rm S}(n)} &=& \tilde{x}_{21}^{{\rm S}(n)}+V_1^{(n)}\tau_c ,
\label{pre-post1}
\\
 {x'}_{32}^{{\rm S}(n)} &=&
     \tilde{x'}_{32}^{{\rm S}(n)}+V_3^{(n)}\tau_c ,
\label{pre-post2}
\end{eqnarray}
where $V_1^{(n)}$ is the relative velocity of the center of mass of the
particles 2 and 3 with respect to the particle 1, and $V_3^{(n)}$ is the
relative velocity of the particle 3 with respect to the center of mass
of the particles 1 and 2,
\begin{eqnarray}
V_1^{(n)} &\equiv& \frac{v_3^{(n)}+v_2^{(n)}}{2}-v_1^{(n)}=v_{21}^{(n)}+\frac{1}{2}v_{32}^{(n)},
\label{V1}
\\
V_3^{(n)} &\equiv& v_3^{(n)}-\frac{v_2^{(n)}+v_1^{(n)}}{2}=v_{32}^{(n)}+\frac{1}{2}v_{21}^{(n)}.
\label{V3}
\end{eqnarray}
The relation Eqs.~(\ref{t1t2}) and (\ref{x21x32}) for the hard particles should be modified for the soft spheres as
\begin{eqnarray}
 t^{{\rm S}(n)}_1 &=& \frac{x^{{\rm S}(n)}_{21}}{|v^{(n)}_{21}|},
\hspace{1cm}
 t^{{\rm S}(n)}_2 = \frac{{x'}^{{\rm S}(n)}_{32}}{|{v'}^{(n)}_{32}|},
\label{t1t2ss}
\end{eqnarray}
and
\begin{eqnarray}
\tilde{x}^{{\rm S}(n)}_{21} &=& {v'}^{(n-1)}_{21} {t}^{{\rm S}(n-1)}_{2},
\hspace{0.5cm}
\tilde{x'}^{{\rm S}(n)}_{32} = v^{(n)}_{32} t^{{\rm S}(n)}_{1},
\label{x21x32ss}
\end{eqnarray}
respectively. Combining Eqs.~(\ref{pre-post1}), (\ref{pre-post2}), (\ref{t1t2ss}) and (\ref{x21x32ss}), we can write
\begin{eqnarray}
t_1^{{\rm S}(n)}
  &=& \frac{{v'}_{21}^{(n-1)}}{|v_{21}^{(n)}|}\frac{v_{32}^{(n-1)}}{|{v'}_{32}^{(n-1)}|}t_1^{{\rm S}(n-1)}
\nonumber\\
 &+&
 \frac{1}{|v_{21}^{(n)}|}\left[\frac{{v'}_{21}^{(n-1)}}{|{v'}_{32}^{(n-1)}|}V_3^{(n-1)}
 +V_1^{(n)}\right]\tau_c.
\label{t1s_1}
\end{eqnarray}
Substituting Eqs.~(\ref{V1}) and (\ref{V3}) into Eq.~(\ref{t1s_1}) and using the ratio of the relative velocities $m^{(n)}$ and
${m'}^{(n)}$ defined in Appendix A, $t_1^{{\rm S}(n)}+\tau_c$ can be expressed as
\begin{eqnarray}
t_1^{{\rm S}(n)} + \tau_c
 &=&
 \frac{1}{e^2}\frac{m^{(n)}m^{(n-1)}}{({m'}^{(n-1)})^2} \left(t_1^{{\rm S}(n-1)}+\tau_c\right)
\nonumber\\
 &+&
 \frac{1}{2}m^{(n)}
 \left[
  \frac{1}{(e {m'}^{(n-1)})^2} -1
  \right] \tau_c .
\label{t1s_2}
\end{eqnarray}
Note that the sequence of the particle velocities $v_{i}^{(n)} (i=1,2,3)$ and that of $m^{(n)}$ and ${m'}^{(n)}$ are
completely determined by the collision laws and their initial conditions regardless whether the particles are hard or soft.
Using the relations Eqs.~(\ref{mmdash}) and (\ref{mdashm}) and the fact $m^{(n)},\,{m'}^{(n)} < 0$ in the collapse
like collision processes, it can be shown that the second term on the right-hand side of Eq.~(\ref{t1s_2}) is negative.

If $m^{(n)}$ and ${m'}^{(n)}$ converge sufficiently fast to their stable fixed point values $m^{*}$ and ${m'}^{*}$,
and we start to count the number of collisions after
$m^{(n)}\approx m^{*}$ and ${m'}^{(n)}\approx {m'}^{*}$ are reached,
we can write
\begin{eqnarray}
\lefteqn{
t_1^{{\rm S}(n)} + \tau_c
}
\nonumber\\
&\approx&
q \left(t_1^{{\rm S}(n-1)}+\tau_c\right)+\frac{m^{*}}{2}\left[\frac{1}{(e {m'}^{*})^2}-1\right]\tau_c ,
\label{t1s_3}
\end{eqnarray}
for any $n\ge 1$.

Because of symmetry with regard to exchange of particles, $t_2^{{\rm S}(n)}$ should have the same expression as Eq.~(\ref{t1s_3}).
Substituting Eq.~(\ref{t1s_3}) into Eq.~(\ref{t12s}), we finally obtain Eqs.~(\ref{t12ss}) and (\ref{deltat12ss}):
\begin{eqnarray}
 t_{12}^{{\rm S}(n)}=q t_{12}^{{\rm S}(n-1)}-f\tau_c,
\end{eqnarray}
where
\begin{eqnarray}
 f\equiv -m^{*}\left[\frac{1}{(e {m'}^{*})^2}-1\right],
\end{eqnarray}
which is a positive function of $e$.


\end{document}